\documentclass[amsmath,amssymb,showpacs,draft,floatfix]{revtex4}
\usepackage{graphicx}
\usepackage{latexsym}
\begin{document}

\title{\textbf{Generalized qubits of the vibrational motion of a trapped ion}}
\author{L.M. Ar\'evalo Aguilar$^1$ and H. Moya-Cessa$^2$}
\date{\today}
\address{$^1$Centro de Investigaciones en Optica, A.C.,
Prolongaci\'on de Constituci\'on No. 607,Apartado Postal No. 507,
Fracc. Reserva Loma Bonita, 20200 Aguascalientes, Ags., Mexico \\
$^2$Instituto Nacional de Astrof\'{\i}sica, Optica y Electr\'onica, Apdo. Postal 51 y
216, 72000 Puebla, Pue., Mexico}

\begin{abstract}
We present a method to generate qubits of the vibrational motion of
an ion. The method is developed in the non-rotating wave approximation regime,
therefore we consider regimes where the dynamics has not been studied.
Because the solutions are valid for a more extended range of parameters
we call them generalized qubits.
\end{abstract}
\maketitle

Nonclassical states of the center-of-mass motion of a trapped ion have
played an important role because of the potential practical applications such as
precision spectroscopy \cite{Wine} quantum computation \cite{CirZol, Dan}
and because of fundamental problems in quantum mechanics.
Ways of producing Scrh\"odinger cat states \cite{monroe},
squeezed states \cite{Win1}, nonlinear coherent states \cite{kis,mato}, number states,
specific superpositions of them, and in particular, robust to noise (spontaneous
emmision) qubits have been proposed \cite{moyac}. In theoretical and
experimental studies of a laser interacting with a single trapped ion it has
been usually considered the case in which it may be modeled as a
Jaynes-Cummings interaction \cite{Win1,winl,Blockley}, then exhibiting the
peculiar features of this model like collapses and revivals \cite{Cir1}, and
the generation of nonclassical states common to such a model or
(multi-photon) generalizations of it \cite{Cir2,Cir3,Blatt}. In treating this
system usually two rotating wave approximations (RWA's) are done (the first
related to the laser [optical] frequency and the second to the vibrational
frequency of the ion), to remove counter-propagating terms of the Hamiltonian
(that can not be treated analytically). Approximations on the Lamb-Dickke
parameter, $\eta$, are usually done, considering it much smaller than unity.
Additionaly, other approximations are done, based on the intensity of the
laser shining on the trapped ion: the low-excitation regime $\Omega \ll \nu $
and the strong-excitation regime $\Omega \gg \nu $ \cite{Poyatos}, with $
\Omega$ being the intensity of the field, and $\nu$ the vibrational
frequency of the ion.

Recently  there has been an alternative approach to the
study of this dynamics \cite{Moya1}. In this approach a unitary transformation is used in
order to linearise the ion-laser Hamiltonian. This transformation has been also used to
propose schemes for realising faster logic gates for quantum information processing
\cite{Dan}.
Under this unitary
transformation the Hamiltonian takes exactly the form (note that \textit{not}
an effective form) of the Jaynes-Cummings Hamiltonian plus an extra term (an
atomic driving term). In such a case a RWA may be done \cite{Moya1} in
order to have an analytical solution for this problem, but it brings with
it a new condition: $\Omega$ of the order of $2\nu$ (note however that this is a
new regime) and $\eta$ still much less than unity.
Later, another transformation \cite{Moya2} was used to diagonalize the
linearized ion-laser Hamiltonian, without further conditions on $\Omega $ or
$\eta$. This allowed the diagonalization of the Hamiltonian only in the ion
basis. Exact eigenstates of the ion-laser
Hamiltonian, i.e. trapping states for this system have been found \cite{Moya3},
but because
they do not form a basis, a complete (exact) solution may be found only
for such states (eigenstates).

In this contribution we consider the complete Hamiltonian for the ion-laser
interaction, linearise it as in \cite{Moya1} and further unitarily transform
it, without performing the RWA to obtain an effective Hamiltonian that can
be easily solved . This is an extension of a method of small rotations
recently applied by Klimov and S\'anchez-Soto \cite{Klimov} to the
Dicke model and other systems (they apply small rotations on the atomic
basis). This allows us to obtain a more general solution valid in a more extended
range of parameters. We apply this solution to a simple initial state and show that
qubits may be produced. Because may be produced with less constrains on the
parameters, we call them generalized qubits.

We consider the Hamiltonian that describes the interaction of a single
two-level trapped ion with a laser beam (we set $\hbar=1$) \cite
{winl,Poyatos}
\begin{equation}
\hat{H}= \nu \hat{n} + \frac{\delta}{2} \hat{\sigma}_{z} + \Omega( \hat{
\sigma}_- e^{-i \eta(\hat{a}+\hat{a}^{\dagger})} + \hat{\sigma}_+ e^{i \eta(
\hat{a}+\hat{a}^{\dagger}) }),  \label{INT1}
\end{equation}
where $\hat{n}=\hat{a}^{\dagger}\hat{a}$,  with $\hat{a}^{\dagger} $ ( $\hat{
a}$ ) the ion vibratonial creation (anihilation) operator, and $\hat{\sigma}
_{+} =|e\rangle \langle g| $ ($\hat{\sigma}_{-} =|g\rangle \langle e| $ ) is
the electronic raising (lowering) operator, $|e\rangle$ ($|g\rangle$) denoting the excited
(ground) state of the ion. The detuning $\delta$ is defined as the difference between the atomic
transition frequency ($\omega_0$) and the frequency of laser($\omega_L$).

Applying the unitary transformation $\hat{T}_1 $ \cite{Moya1}
\begin{equation}
\hat{T}_1= \frac{1}{\sqrt{2}} \left( \frac{1}{2}[\hat{D}^{\dagger}(\beta)+
\hat{D}(\beta)]\hat{I} + \frac{1}{2}[\hat{D}^{\dagger}(\beta)-\hat{D}(\beta)]
\hat{\sigma}_z - \hat{D}^{\dagger}(\beta)\sigma_- + \hat{D}(\beta)\sigma_+
\right),
\end{equation}
(where $\hat{I}= |g\rangle \langle g| + |e\rangle \langle e|$ and $\hat{D}
(\beta)$ is the displacement operator with $\beta= \frac{i \eta}{2} $ its
amplitude) to (\ref{INT1}), we obtain the linearised Hamiltonian (in the on resonance case i. e. $
\delta =0$), given by \cite{Moya1}
\begin{equation}
\hat{H}_{1}=\hat{T}_1 \hat{H} \hat{T}^{\dagger}_1 = \nu \hat{n} + \Omega
\hat{\sigma}_{z} + \/
i \frac{\eta \nu}{2} (\hat{a} - \hat{a}^{\dagger})(\hat{
\sigma}_{-}+ \hat{\sigma}_{+}) + \frac{\nu \eta^{2}}{4}.  \label{linearized}
\end{equation}

We now apply the unitary tranformation
\begin{equation}
\hat{T}_2=e^{-i\epsilon (\hat{a}+\hat{a}^{\dagger})(\hat{\sigma}_{+} + \hat{
\sigma}_{-})},  \label{tra2}
\end{equation}
to (\ref{linearized}) to obtain $\hat{H}_2=\hat{T}_2 \hat{H}_1 \hat{T}
^\dagger_2 $,

\begin{eqnarray}
\hat{H}_{2}  & = & \nu \left( \hat{n}+i\epsilon (\hat{a}-\hat{a}^{\dagger})
(\hat{\sigma}_{-}+\hat{\sigma}_{+})+\epsilon ^{2}\right)  \nonumber \\
&  & +\Omega \left( \sigma_{z}\cos [2\epsilon (\hat{a}+\hat{a}^{\dagger})]
+i(\sigma _{-}-\sigma_{+})\sin [2\epsilon (\hat{a}+\hat{a}^{\dagger})]\right)  \nonumber \\
 & & + i\frac{\eta \nu }{
2}(\hat{a}-\hat{a}^{\dagger })(\hat{\sigma}_{-}+\hat{\sigma}_{+})+\frac{\nu
\epsilon \eta ^{2}}{4}.
\label{h2}
\end{eqnarray}
By considering $\epsilon \ll 1$, we can approximate (\ref{h2}) as (we
disregard constant terms that only contribute to a shift of energies)
\begin{eqnarray}
\hat{H}_{2} & \approx & \nu \left( \hat{n}+i\epsilon (\hat{a}-\hat{a}^{\dagger })(
\hat{\sigma}_{-}+\hat{\sigma}_{+})\right)
+\Omega \left( \sigma_{z}+2i\epsilon (\sigma _{-}-\sigma _{+})(\hat{a}+\hat{a}^{\dagger
})\right)  \nonumber \\
   &   & +   i\frac{\eta \nu }{2}(\hat{a}-\hat{a}^{\dagger })(\hat{\sigma}_{-}+\hat{\sigma}_{+}),
\end{eqnarray}
and by setting
\begin{equation}
\epsilon =-\frac{\eta }{2}\frac{\nu }{\nu +2\Omega },
\end{equation}
we finally obtain
\begin{equation}
\hat{H}_{2}=\nu \hat{n}+\Omega \hat{\sigma}_{z}+i\lambda (\hat{\sigma}_{+}
\hat{a}-\hat{a}^{\dagger }\hat{\sigma}_{-}).
\label{JCM}
\end{equation}
Note that the {\it coupling constant}, $\lambda =\frac{2\eta \nu \Omega }{
\nu +2\Omega },$ has changed (before it was $\frac{\eta \nu }{2}$).

We should remark that transformation (\ref{tra2}) requires $\epsilon \ll 1$
and this may be achieved in different forms:

 {\it a)} $\eta \ll 1$ and $\nu$ and $
\Omega$ \textit{any} numbers (this is, it can be $\nu \ll \Omega$, $\nu
\gg \Omega$ or of the same order of magnitude);

{\it b)} no restrictions on $\eta$ and $\nu \ll \Omega$ or,

 {\it c)} $\eta < 1$ and $\nu < \Omega$ (note that, for instance, a value of $\eta=0.3$ and $
\Omega=2\nu$ gives $\epsilon = -0.03$).

The three possibilities above allow the Hamiltonian to be approximated to
first order and to disregard terms of second and higher orders.

We are now in the position to give a solution to the Hamiltonian (\ref{INT1}), that reads
\begin{equation}
|\Psi (t)\rangle= \hat{T}^{\dagger} \hat{U} \hat{T} |\Psi (0) \rangle,
\label{solu}
\end{equation}
where we have written
\begin{eqnarray}
\hat{T} & = & \hat{T}_2 \hat{T}_1   \nonumber \\
& = &  \frac{1}{\sqrt{2}} \left( \frac{1}{2}[\hat{D}
^{\dagger}(\beta_-)+\hat{D}(\beta_-)]\hat{I} + \frac{1}{2}[\hat{D}
^{\dagger}(\beta_-)-\hat{D}(\beta_-)]\hat{\sigma}_z - \hat{D}
^{\dagger}(\beta_-)\sigma_- + \hat{D}(\beta_-)\sigma_+ \right),
\end{eqnarray}
with $\beta_-= i(\frac{\eta}{2}-\epsilon)$ and where $|\Psi (0) \rangle$ is
the initial wave function and $\hat{U} $ is the evolution operator of the
off-resonant Jaynes-Cummings Model
\begin{equation}
\hat{U}=e^{-it(\nu\hat{n}+\frac{1}{2}\nu\hat{\sigma}_{z})} e^{-it[\Delta\hat{
\sigma}_{z}+i\lambda( \hat{a}\hat{\sigma}_{+}-\hat{a}^{\dagger}\hat{\sigma}
_{-})]},  \label{unitario}
\end{equation}
where we defined $\Delta=\Omega-\frac{\nu}{2}$. Equation (\ref{unitario})
may be finally written in the form \cite{Sten}
\begin{equation}
\hat{U}= e^{-it(\nu\hat{n}+\frac{1}{2}\nu\hat{\sigma}_{z})}
\left( \frac{1}{2}[\hat{U}_{11}+\hat{U}_{22}]\hat{I} + \frac{1}{2}[
\hat{U}_{11}-\hat{U}_{22}]\hat{\sigma}_z + \hat{U}_{21}\sigma_- +
\hat{U}_{12}\sigma_+ \right),
\end{equation}
with
\begin{equation}
\hat{U}_{11} = \cos\hat{\alpha}_{\hat{n}+1}t-i\Delta \frac{\sin\hat{\alpha}_{
\hat{n}+1}t}{\hat{\alpha}_{\hat{n}+1}},
\end{equation}
\begin{equation}
\hat{U}_{12} = \lambda\hat{a} \frac{\sin\hat{\alpha}_{\hat{n}}t}{\hat{\alpha
}_{\hat{n}}},
\end{equation}
\begin{equation}
\hat{U}_{21} = -\lambda\hat{a}^{\dagger} \frac{\sin\hat{\alpha}_{\hat{n}+1}t}
{\hat{\alpha}_{\hat{n}+1}},
\end{equation}
and
\begin{equation}
\hat{U}_{22} = \cos\hat{\alpha}_{\hat{n}}t+i\Delta \frac{\sin\hat{\alpha}_{
\hat{n}}t}{\hat{\alpha}_{\hat{n}}},
\end{equation}
where
\begin{equation}
\hat{\alpha}_{\hat{n}} =\sqrt{\Delta^2+\lambda^2\hat{n} }.
\end{equation}
By considering the ion initially in its excited state and in a coherent
(vibrational) state with amplitude $\beta_-$, i.e.
\begin{equation}
|\Psi (0) \rangle= |e\rangle |\beta_- \rangle ,
\end{equation}
we obtain the evolved wave function at time $t$
\begin{eqnarray}
|\Psi (t) \rangle & = & \frac{1}{2}
\hat{D}(\beta_-)\left([e^{-i\frac{\nu t}{2}}(\cos\alpha_1t-i\frac{\Delta}{\alpha_1}\sin\alpha_1t)
+e^{i\Omega t}]|0 \rangle +  \frac{\lambda}{\alpha_1}e^{-i\frac{\nu t}{2}}\sin\alpha_1 t |1\rangle
\right) |e \rangle \nonumber \\
  & + &\frac{1}{2}
\hat{D}^{\dagger}(\beta_-)\left([e^{-i\frac{\nu t}{2}}(\cos\alpha_1t-i\frac{\Delta}{\alpha_1}\sin\alpha_1t)
-e^{i\Omega t}]|0 \rangle - \frac{\lambda}{\alpha_1}e^{-i\frac{\nu t}{2}}\sin\alpha_1 t |1\rangle
\right) |g\rangle
\end{eqnarray}
with $\alpha_1 = \langle 1|\hat{\alpha}_{\hat{n}}  |1 \rangle$.
Note that $\hat{D}(\beta_-)|k \rangle = |\beta_-,k\rangle$ is a displaced number state
\cite{Oli}. By measuring either the excited state or ground state of the ion (there are standard
techniques to do so, see for instance \cite{winl}),
we end up with a superposition of displaced number states with amplitude $\beta_-$ or
$-\beta_-$, respectively. Therefore, by displacing the resulting state if
the ion is measured in the excited state by $-\beta_-$ (or by $\beta_-$ if measured
in the ground state, we produce a qubit of the vibrational motion of an ion. To clarify this
point, let us assume the ion is measured in the excited state. The vibrational wave
function collapses to the state

\begin{equation}
|\Psi_{vib} (t) \rangle = \frac{1}{N}
\hat{D}(\beta_-)\left([e^{-i\frac{\nu t}{2}}(\cos\alpha_1t-i\frac{\Delta}{\alpha_1}\sin\alpha_1t)
+e^{i\Omega t}]|0 \rangle +  \frac{\lambda}{\alpha_1}e^{-i\frac{\nu t}{2}}\sin\alpha_1 t |1\rangle
\right) ,
\end{equation}
where $N$ is the normalization constant. By finally displacing the state by an amplitude
$-\beta_-$, we obtain the (qubit) state
\begin{equation}
|\Psi_{d} (t) \rangle = \frac{1}{N}
\left([e^{-i\frac{\nu t}{2}}(\cos\alpha_1t-i\frac{\Delta}{\alpha_1}\sin\alpha_1t)
+e^{i\Omega t}]|0 \rangle +  \frac{\lambda}{\alpha_1}e^{-i\frac{\nu t}{2}}\sin\alpha_1 t |1\rangle
\right).
\end{equation}

Moreover, without (conditional) measurement, and by controling the interaction time between the
ion and the laser beam, setting it to $\alpha_1 t= \pi$, the (Schr\"odinger cat) state
\begin{equation}
|\Psi(\alpha_1 t= \pi)\rangle = \frac{1}{2}\left(
(e^{i\Omega t}-e^{-i\nu t})|\beta_- \rangle |e\rangle -
(e^{i\Omega t}+e^{-i\nu t})|-\beta_- \rangle |g\rangle \right),
\end{equation}
is generated. This state was produced in \cite{monroe} and also studied in
\cite{wal}. Note that the quantity $\beta_-$  is slightly larger than the Lamb-Dicke
parameter, which in general is not needed to be small.

In this contribution we have suceeded in solving the problem of trapped ion
interacting resonantly ($\delta=0$ case)  with a laser field whitout making
the rotating wave aproximation, but applying a method of small rotations
that allowed such solution. It should be remarked that now there are new
regimes that can be reached, since the usual approximations as the low intensity or
high intensity regimes were not considered.

We thank CONACYT (Consejo Nacional de Ciencia y Tecnolog\'{\i}a) for support and the referee
for useful comments.

\end{document}